\documentclass[fleqn,usenatbib]{mnras}

\usepackage{longtable}
\usepackage{amssymb}
\usepackage{amsmath}

\usepackage[T1]{fontenc}

\DeclareRobustCommand{\VAN}[3]{#2}
\let\VANthebibliography\thebibliography
\def\thebibliography{\DeclareRobustCommand{\VAN}[3]{##3}\VANthebibliography}

\usepackage{graphicx}	
\usepackage{amsmath}	
\usepackage{amssymb}	

\usepackage{newtxtext,newtxmath}

\newcommand{\Msun}{\mbox{M$_{\odot}$}}

\newcommand{\lppr}{\stackrel{<}{\scriptstyle \sim}}
\newcommand{\lappr}{\raisebox{-0.4ex}{$\lppr$}}
\newcommand{\gppr}{\stackrel{>}{\scriptstyle \sim}}
\newcommand{\gappr}{\raisebox{-0.4ex}{$\gppr$}}

\newcommand{\mesa}{{\sc mesa}}
\newcommand{\bse}{{\sc bse}}

\title[Magnetic white dwarfs in double degenerates]{Magnetic dynamos in white dwarfs --III: 
explaining the occurrence of strong magnetic fields in close double white dwarfs}

\author[M.R.Schreiber et al.]{
Matthias.R. Schreiber,$^{1,2}$\thanks{E-mail: matthias.schreiber@usm.cl}
Diogo Belloni,$^{1}$
Monica Zorotovic,$^{3}$
Sarai Zapata,$^{1}$
\newauthor
Boris T. G\"ansicke,$^{4}$
Steven G. Parsons$^{5}$
\\
$^{1}$Departamento de F{\'i}sica, Universidad T\'ecnica Federico Santa Mar\'ia, Av. España 1680, Valpara{\'i}so, Chile\\
$^{2}$Millennium Nucleus for Planet Formation, NPF, Valpara{\'i}so, Chile\\
$^{3}$Instituto de F\'isica y Astronom\'ia, Universidad de Valpara\'iso, Av. Gran Breta\~na 1111, Valpara\'iso, Chile\\
$^{4}$Department of Physics, University of Warwick, Coventry CV4 7AL, UK\\
$^{5}$ Department of Physics and Astronomy, University of Sheffield, Sheffield S3 7RH, UK\\
}

\date{Accepted XXX. Received YYY; in original form ZZZ}

\pubyear{2021}

\begin{document}
\label{firstpage}
\pagerange{\pageref{firstpage}--\pageref{lastpage}}
\maketitle

\begin{abstract}
The origin of strong ($\gappr1MG$) magnetic fields in white dwarfs has been a puzzle for decades. Recently, a dynamo mechanism operating in rapidly rotating and crystallizing white dwarfs has been suggested to explain the occurrence rates of strong magnetic fields in white dwarfs with close low-mass main sequence star companions. 
Here we investigate whether the same mechanism may produce strong magnetic fields in close double white dwarfs. 
The only known strongly magnetic white dwarf that is part of a close double white dwarf system, the magnetic component of NLTT\,12758, is rapidly rotating and likely crystallizing and therefore the proposed dynamo mechanism represents an excellent scenario for the origin of its magnetic field. Presenting a revised formation scenario for NLTT\,12758, we find a natural explanation for the rapid rotation of the magnetic component. 
We furthermore show that it is not surprising that strong magnetic fields have not been detected in all other known double white dwarfs.
We therefore conclude that the incidence of magnetic fields in close double white dwarfs supports the idea that a rotation and crystallization driven dynamo plays a major role in the generation of strong magnetic fields in white dwarfs. 
\end{abstract}




\section{Introduction}

White dwarfs have been speculated to potentially have strong magnetic fields (exceeding 1 MG) since 1947 \citep{blackett47}, but the first detection of a magnetic field in a white dwarf was only obtained more than twenty years later \citep[][]{kempetal70-1}. 
Ever since, the question why some white dwarfs become strongly magnetic while others do not, has been one of the fundamental unsolved issues of stellar evolution. 
Answering this question appears to be very complicated largely because strongly magnetic white dwarfs are found in different relative numbers among single white dwarfs, white dwarfs in close detached binaries, and white dwarfs in close semi-detached binary stars.  

The incidence of strongly magnetic white dwarfs in magnitude-limited surveys of isolated white dwarfs is roughly five per cent \citep{kepleretal13-1}, but the fraction is clearly higher among nearby white dwarfs which contain more old systems \citep{kawkaetal07-1,bagnulo+landstreet20-1,bagnulo+landstreet21-1} which indicates that the fractions derived from magnitude limited surveys potentially largely underestimate the true fraction of strongly magnetic white dwarfs.  

Among close detached white dwarf binaries the fraction of systems containing a strongly magnetic white dwarf
is very small, at most a few percent, and all known systems are close to Roche-lobe filling and contain cold white dwarfs with temperatures $\lappr10\,000$\,K \citep{parsonsetal21-1}. In contrast, the fraction of strongly magnetic white dwarfs among their semi-detached descendants is very high, exceeding most likely one third \citep{palaetal20-1}. 

Present versions of the models suggested for generating strong magnetic fields in white dwarfs, the fossil field \citep[e.g.][]{toutetal04-1,braithwaiteetal04-1}, the double degenerate merger \citep{garcia-berroetal12-1}, and the common envelope dynamo scenarios \citep[e.g.][]{toutetal08-1}, cannot reproduce these observations. 
Either the predicted numbers of strongly magnetic white dwarfs are too small as in the fossil field scenario \citep{kawka+vennes04-1,toutetal04-1,wickramasinghe+ferrario05-1,auriereetal07-1}, or the number of predicted strongly magnetic white dwarfs is far too large as in the common envelope scenario \citep{belloni+schreiber20-1}, or fails to reproduce the large number of magnetic white dwarfs in interacting binary stars as in the double degenerate merger scenario. In addition, in their current form, none of the above scenarios can explain the absence of strongly magnetic white dwarfs in young detached white dwarf/main sequence binary stars
\citep{liebertetal05-1,liebertetal15-1,belloni+schreiber20-1,schreiberetal21-1}. 

We have recently developed a new model for the generation of magnetic fields in white dwarfs that can potentially explain the different occurrence rates of strongly magnetic white dwarfs with main sequence star companions \citep{schreiberetal21-1}. 
This scenario is based on the idea first put forward by \citet{isernetal17-1} who proposed that a crystallization and rotation driven dynamo similar to those operating in planets and low-mass stars could be generating magnetic fields in white dwarfs. 

According to the scenario developed by \citet{schreiberetal21-1}, white dwarfs in close binaries emerge from common envelope evolution and evolve towards shorter orbital periods without hosting a strong magnetic field. 
As soon as mass transfer starts these post common envelope binaries become cataclysmic variables (CVs) and the white dwarf accretes mass and angular momentum. If the spun-up white dwarf has cooled enough to contain a crystallizing core, its structure consists of an outer carbon rich convective zone and a solid inner oxygen rich nucleus \citep{isernetal17-1,schreiberetal21-1}. 
This situation resembles that of the interior of planets and low-mass stars and the convective motion in the liquid mantle can generate a magnetic field.

\citet{isernetal17-1} estimated the field strength generated by this dynamo based on scaling laws that are used for planets and low-mass stars \citep{christensen+aubert06-1,christensenetal09-1} and derived an upper limit of around one MG. However, existing scaling laws are most likely not appropriate for white dwarfs as the much higher magnetic Prandtl number in white dwarfs could imply a much larger field strength \citep{brandenburg14-1,bovinoetal13-1}. Very recently, \citet{ginzburgetal22-1} argued that the convective turn-over time is much larger than assumed by \citet{isernetal17-1} which would imply that white dwarfs are almost always in the fast rotation regime. Combining this finding with a scaling law based on the balance between the Lorentz
and Coriolis forces, even permits the generation of strong magnetic fields (reaching 100 MG) without postulating a magnetic field enhancement due to the white dwarf’s Prandtl number. 

If indeed strong magnetic fields are generated in the white dwarfs in CVs due to this dynamo, these fields may connect with the field of the secondary star and synchronizing torques transfer spin angular momentum from the white dwarf to the orbital motion which can cause the system to detach. At first the detached system will appear for a short period of time as a radio pulsing detached white dwarf binary similar to 
AR\,Sco \citep{marshetal16-1}, then evolve into a synchronized detached pre-polar \citep{schwopeetal09-1,parsonsetal21-1} and subsequently into a semi-detached magnetic CV (polar). 

In addition to offering an explanation for the origin and evolution of magnetic white dwarfs in close binary stars with a main sequence star companion, we could recently show that the dynamo also naturally explains the low occurrence rate of high accretion rate magnetic CVs in globular clusters \citep{bellonietal21-1} and the increased incidence of magnetism among cold metal polluted white dwarfs \citep{schreiberetal21-2}. 

Despite this success, we note that some individual systems containing magnetic white dwarfs require an alternative mechanism for the magnetic field generation. The recently discovered post common envelope binary CC\,Cet \citep{wilsonetal21-1} and the polar EY\,Eri \citep{beuermannetal20-1} both contain
most likely magnetic helium core white dwarfs and one of the main ingredients of the dynamo scenario is a crystallizing core consisting of carbon and oxygen. In addition, the potentially weakly magnetic white dwarf in the post common envelope binary V471\,Tau \citep{sionetal12-1} is too hot to be crystallizing. 

Given this situation, it is of utmost importance to further investigate to which degree the dynamo scenario may solve the long standing puzzle of the origin of strong magnetic fields in white dwarfs and in which cases an alternative mechanism is required. 
Here we apply the suggested dynamo mechanism to close double white dwarf binaries and investigate whether it can explain the occurrence rate and characteristics of strongly magnetic white dwarfs in these systems.

\section{Known double white dwarfs and magnetic fields}

In order to investigate whether the new scenario for magnetic field generation may explain the observations of close (here defined as an orbital period below 35 days) double white dwarfs, we compiled a list of the currently known systems (see Appendix A, Table\,\ref{tab:dwds}). 
At present, only one system is known to host a strongly magnetized white dwarf.
NLTT\,12758 consists of a magnetic white dwarf with a field strength of about 3.1 MG and an apparently non-magnetic white dwarf companion. The non-magnetic H-rich (DA) white dwarf is more massive ($M=0.83\pm0.03 \Msun$) than its 
magnetic (DAP) companion ($M=0.69\pm0.05\Msun$), the orbital period of the binary is 1.154 days, the spin period of the magnetic white dwarf has been measured to be 23 minutes, and the cooling ages of both white dwarfs are comparable 
\citep{kawkaetal17-1}.  

The recently proposed magnetic dynamo requires relatively rapid rotation and a crystallizing core 
to generate strong magnetic fields in white dwarfs. Saturation of the dynamo is assumed to occur for
spin periods of just a few seconds \citep[see][for more details]{isernetal17-1,schreiberetal21-1}. 
The magnetic and lower mass white dwarf in NLTT\,12758 is rapidly rotating with a spin period of 23 minutes
which is much shorter than the typical spin period of non-magnetic single white dwarfs \citep{hermesetal17-1}
but longer than the short periods of the order of seconds proposed for white dwarfs in CVs \citep{schreiberetal21-1}. 
One would therefore expect the magnetic field strength of the magnetic white dwarf in NLTT\,12758
to be smaller than the strongest magnetic fields observed in CVs. 
Indeed, in magnetic CVs the field strengths of the white dwarfs reaches several hundred MG with an average field strength of $\sim$30\,MG \citep{ferrarioetal15-1,Belloni2020}
and the average field strength of pre-polars is $\sim$60-70\,MG \citep{parsonsetal21-1} 
while for the magnetic component in NLTT\,12758 a field strength of just 3.1 MG has been measured \citep{kawkaetal17-1}. 

\begin{figure*}
    \centering
    \includegraphics[height=9cm]{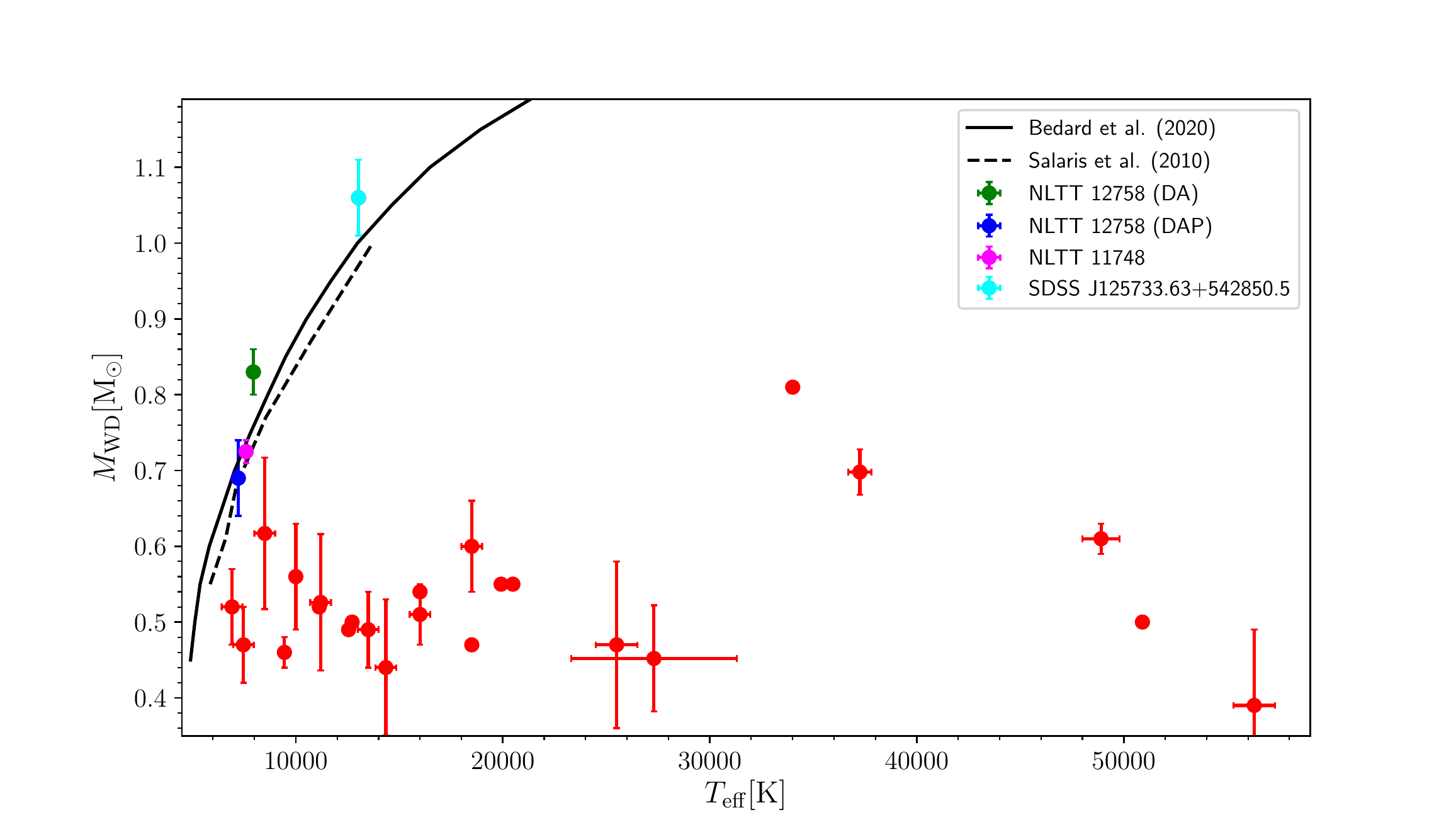}
    \caption{Comparison of the mass and effective temperature of the known white dwarfs that form part of a close (period less than 35 days) double white dwarf (from Table\,\ref{tab:dwds}) with the onset of crystallization according to the cooling sequences by \citet[][black dashed line]{salarisetal10-1} and \citet[][black solid line]{bedardetal20-1}. The vast majority of the white dwarfs are too hot to be crystallizing. 
    Among the few exceptions is the magnetic component in NLTT\,12758. For this white dwarf the measured parameters are consistent with crystallization in its core. As in addition the spin period has been measured to be very short (23 min.) it appears plausible that the rotation and crystallization driven dynamo generated the strong magnetic field. Three more white dwarfs are also consistent with having a crystallizing core. This might indicate that those white dwarfs, in contrast to the magnetic component in NLTT\,12758, did not accrete angular momentum during their evolutionary history or that their magnetic field has not been detected yet. } \label{fig:M-Teff}
\end{figure*}

The second criterion for the dynamo to work is that the core of the white dwarf must have started to crystallize.  
The onset of crystallization depends on the temperature and the white dwarf mass. The currently available cooling tracks that include crystallization differ slightly in the predicted temperatures for the onset of crystallization. 
In Fig.\ref{fig:M-Teff} we compare the 
onset of crystallization from the models by \citet{salarisetal10-1} and \citet{bedardetal20-1} with the position 
of all C/O white dwarfs in close double white dwarfs with known mass and effective temperature. 

The measured parameters of the magnetic white dwarf component in NLTT\,12758 (highlighted in blue in Fig.\,\ref{fig:M-Teff}) are consistent with a  
crystallizing core in the white dwarf and therefore the rotation and crystallization driven dynamo offers a plausible explanation for its magnetic nature. 
Most other white dwarfs that are members of close double white dwarfs are either He-core white dwarfs 
for which the dynamo scenario does not apply and/or are too hot for having crystallizing cores which is also consistent with the 
dynamo scenario for magnetic field generation in white dwarfs. 
However, of the few double white dwarfs containing 
C/O white dwarfs, three additional ones have sufficiently cooled to be crystallizing. 

Apart from the DAP white dwarf in NLTT\,12758, 
components close to or beyond the onset of crystallization 
are its higher mass DA companion and the more massive white dwarfs in SDSS\,J125733.63+542850.5 
(hereafter SDSS\,J1257+5428) and NLTT\,11748 (the white dwarfs highlighted with cyan and magenta in Fig.\,\ref{fig:M-Teff}). 
This leads to an important follow-up question: {\em{why is NLTT\,12758 
the only close double white dwarf where a strongly magnetic component has been detected? }}
If our hypothesis that the rotation and crystallization driven dynamo generated the magnetic field in 
NLTT\,12758 is correct, 
the other white dwarfs with crystallizing cores must either not have sufficiently spun up during their formation 
or we simply have not been able to detect their strong magnetic fields. 
To understand magnetic field generation in close double white dwarfs we therefore need to investigate the evolutionary history of 
NLTT\,12758 and consider potential evolutionary differences to NLTT\,11748 and SDSS\,J1257+5482 
as well as possible observational biases. 
We start by taking a closer look at the formation of NLTT\,12758.

\section{Was NLTT~12758 formed through two common envelope phases?}

Most close white dwarf binary stars form through common envelope evolution \citep{webbink84-1,zorotovicetal10-1,ivanovaetal13-1} which occurs when the more massive star 
fills its Roche-lobe on the first giant branch (FGB) or the asymptotic giant branch (AGB). The resulting mass transfer is typically dynamically unstable and leads to the formation of a gaseous envelope around the secondary star and the core of the giant. This envelope is expelled at the expense of orbital energy and orbital angular momentum. Therefore, the emerging post common 
envelope binary stars have typically short orbital periods between a few hours and a few days \citep{nebotetal11-1}.

Common envelope evolution has been shown to successfully explain the observed close white dwarf plus M dwarf binary population \citep{zorotovicetal10-1} and seems to also reproduce the short periods found among some close white dwarf plus FGK secondary star binary systems \citep{hernandezetal21-1}. Two consecutive common envelopes have been suggested as the formation scenario for NLTT\,12758 \citep{kawkaetal17-1} and we therefore start our discussion with a review of this classical scenario for the formation of close double white dwarfs. 

The change in orbital separation during common envelope evolution can be calculated using a simple energy conservation equation that relates the loss of orbital energy of the binary and the binding energy of the envelope. 
This energy conservation equation is typically parameterized with 
the common envelope efficiency $\alpha_{\mathrm{CE}}$ and the binding energy parameter $\lambda$ \citep[e.g.][]{hanetal95-1,dewi+tauris00-1,zorotovicetal10-1}.

While it is clear that the common envelope efficiency $\alpha_{\mathrm{CE}}$ should be between zero and one -- there are strong indications for a relatively low value between $0.2-0.3$ \citep{zorotovicetal10-1} -- there is no agreement
in the literature on the most realistic value of the binding energy parameter $\lambda$ mostly because it remains unclear 
how efficient recombination energy 
can contribute to the process of expelling a common envelope 
\citep{webbink08-1,ivanovaetal15-1,ivanova18-1,sabachetal17-1,gricheneretal18-1,sokeretal18-1}. 

Assuming two common envelope phases and using the \bse\, code, \citet{kawkaetal17-1}  found that the history of NLTT\,12758 can be reproduced assuming an initial binary with stellar masses of 3.75\,\Msun\, and 2.88\,\Msun\, and an orbital period of 2656\,d. 
We could reproduce their result only by assuming that $\sim75$ per cent of the available recombination energy contributed to expelling the envelope.
This assumption can be considered unrealistic as even work that emphasize the potential importance of recombination energy 
exclude such large values \citep{ivanovaetal15-1}.
We believe that this large fraction of recombination energy is related to a misinterpretation of the input parameters of {\bse} (see Appendix B, Fig.\ref{fig:NLT015}, for more details).  

In order to investigate whether more realistic parameter combinations allow to reproduce the formation of NLTT\,12758 through two common envelope phases we reconstructed both potential common envelope phases following \citet{zorotovicetal10-1} and \citet{zorotovicetal14-1}.  
We found that without contributions from recombination energy the system cannot have formed by two consecutive common envelope phases.
Assuming a relatively small fraction of recombination energy ($\leq10$ per cent, but at least $2$ per cent) and a value of the common envelope efficiency exceeding 0.5, reasonable solutions exist. 

These possibilities exist because the $\lambda$ parameter is extremely sensitive to the inclusion of recombination energy.
Even for an efficiency of recombination energy of just $2$ per cent, the $\lambda$ parameter for the more massive white dwarf progenitor (first common envelope phase) is $\gappr3$, leading to a much smaller reduction of the orbital separation with respect to simulations that do not consider recombination energy at all (where $\lambda$ is closer to 1).  

This situation is similar to that of the two white dwarf plus main-sequence post common envelope binaries known so far that require extra energy sources in order to explain their current orbital periods (IK\,Peg and KOI\,3278). However, both systems can be reproduced by assuming that $\lappr2$ per cent of the recombination energy contributed to the ejection process \citep{zorotovicetal14-1} and for NLTT\,12758 we need a slightly larger fraction.
We therefore conclude that two consecutive common envelope phases remain a possible scenario for the evolutionary history of NLTT\,12758 but that an (uncertain and unusual) source of additional energy (possibly recombination energy) is required.  

If two common envelope phases produced NLTT\,12758, the white dwarf with the smaller mass must have formed through the second common envelope phase and the more massive white dwarf must have formed through the first one. This is because the onset of common envelope evolution is defined by dynamically unstable mass transfer which quickly reaches mass transfer time 
scales that are orders of magnitude shorter than the thermal time scale of the secondary star. The secondary star is therefore supposed to accrete only a small amount of mass (according to \citealt{chamandyetal18-1} the total accreted mass is roughly in the range of $10^{-2}-10^{-4}$\,\Msun). As in addition the spiral-in process significantly reduces the orbital separation, the initially less massive star can not evolve further up the giant branches than the first one. In other words, in two consecutive common envelope phases the more massive white dwarf is formed first.  
For NLTT\,12758 this implies that the fast spinning and strongly magnetic white dwarf formed in the second common envelope phase. If that was indeed the evolutionary history of NLTT\,12758, it remains an open question why the magnetic white dwarf is rapidly rotating.

However, the fact that the cooling ages of both white dwarfs in NLTT\,12758 are very similar, i.e. $2.2\pm0.2$ and $1.9\pm0.4$\,Gyr \citep{kawkaetal17-1}
might indicate that the formation history of the system was actually different. If the two progenitor stars evolved off the main sequence at about the same time, their masses must have been similar and therefore the first mass transfer phase could have been stable. 

In fact, according to Table\,5 in \citet{kawkaetal17-1}, even in their solution for two common envelopes the first mass transfer started when both stars had virtually identical masses which, according to \citet{geetal20-1}, should lead to stable mass transfer instead of common envelope evolution if the mass transfer is not fully conservative (i.e. a fraction of the transferred mass leaves the system). 

In the following section, we therefore discuss an alternative formation scenario for NLTT\,12758 which includes stable non-conservative mass transfer and show that it offers a natural explanation for the observed binary and stellar parameters of the system as well as the short spin period of the magnetic white dwarf. 

\section{An alternative formation model: Combining stable mass transfer and common envelope evolution} 

As noted first by \citet{nelemansetal2000-1} two common envelope phases using the classical energy budget equation fail to explain several observed double white dwarf systems. To solve this issue, they suggested the $\gamma$ algorithm based on angular momentum conservation which allows for common envelope evolution without spiral in during the first mass transfer phase. However, by implicitly assuming energy conservation the $\gamma$ algorithm is rather hiding the energy problem instead of solving it \citep{ivanovaetal13-1}. 
As an alternative, \citet{webbink08-1} suggested that instead the first mass transfer could be stable and non-conservative which  
occurs when the mass ratio of the two progenitor stars is close to one \citep{geetal20-1}. Detailed models
for double He-core white dwarfs were subsequently performed and showed that stable non-conservative mass transfer indeed offers an explanation \citep[][]{woodsetal12-1}.
We here explore the evolutionary scenario in which the first white dwarf forms through stable mass transfer for NLTT\,12758 using detailed stellar evolution calculations.  

\subsection{Simulating the evolution with MESA}

We performed our simulations of the binaries with the one-dimensional stellar and binary evolution code Modules for Experiments in Stellar Astrophysics \citep[\mesa][r15140]{Paxton2011, Paxton2013, Paxton2015, Paxton2018, Paxton2019}.

\begin{figure*}
\includegraphics[width=0.7\textwidth]{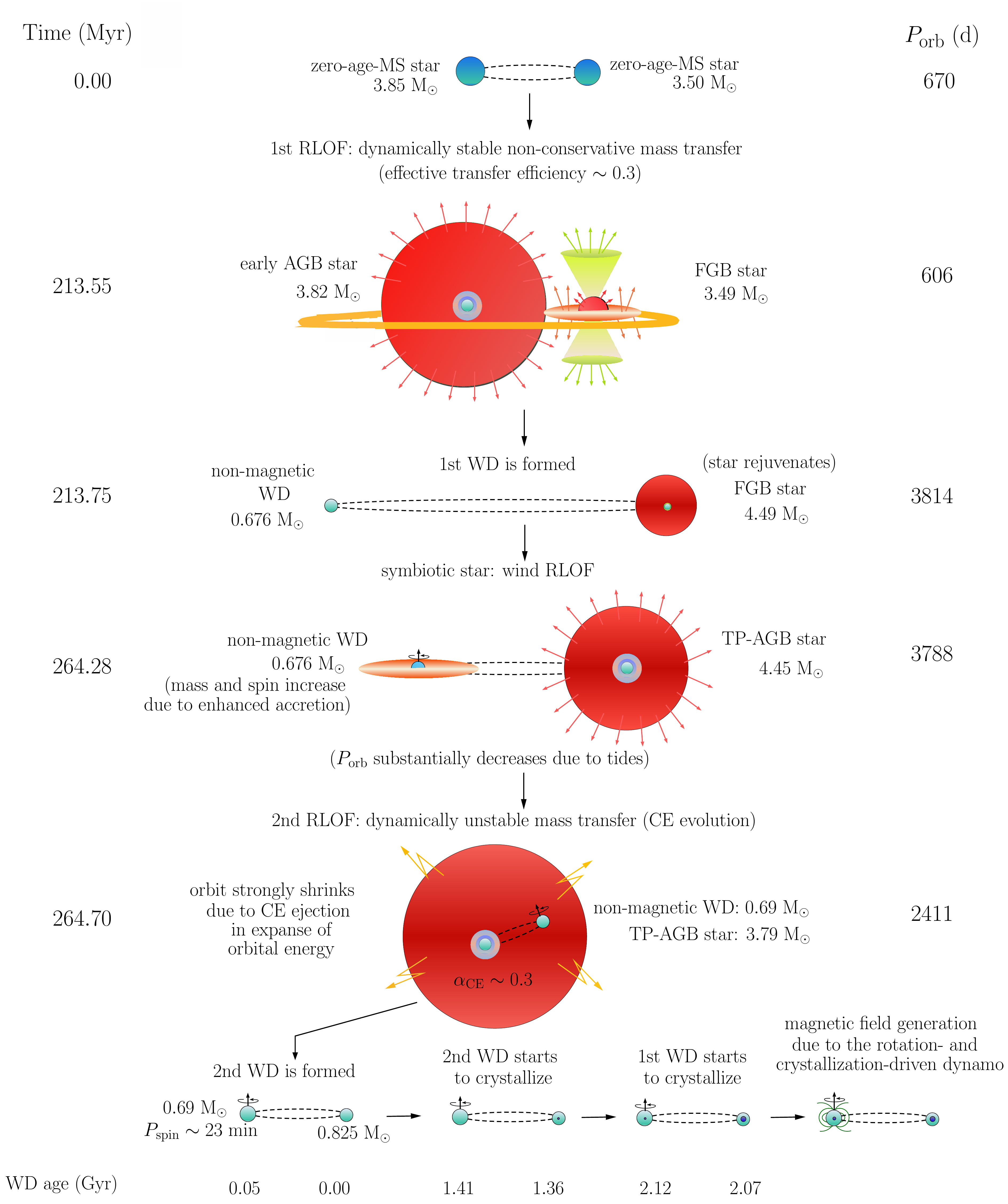}
\caption{The revised evolutionary scenario for NLTT\,12758 consists of stable mass transfer followed by a symbiotic phase and common envelope evolution after which the orbital period is similar to the one we observe today. According to this scenario, the less massive white dwarf forms first and accretes mass and angular momentum during the second mass transfer phase. When this fast spinning white dwarf starts to crystallize, the dynamo mechanism suggested by \citet{schreiberetal21-1} generates the strong magnetic field that has been detected by \citet{kawkaetal17-1}.} 
\label{fig:NEW}
\end{figure*}

%
We accounted for mass loss through winds using the standard MESA implementations.
In particular, for red giants on the FGB, we adopted the \citet{Reimers1975} prescription, assuming a wind efficiency of ${\eta=0.5}$, which is consistent with metallicity-independent estimates using star clusters \citep{McDonald2015}.
For AGB stars, we adopted the \citet{Bloecker1995} recipe, assuming a wind efficiency of ${\eta=0.02}$, which is consistent with the calibration performed by \citet{Ventura2000} using the luminosity function of lithium-rich stars in the Magellanic Clouds.

%
The Roche lobe radius of each star was computed using the fit of \citet{Eggleton1983} and mass transfer rates during Roche lobe overflow (RLOF) are determined following the prescription of \citet{Ritter1988}.
Regarding wind accretion, we assumed two different prescriptions depending on the donor stars.
For FGB stars, we adopted the Bondi-Hoyle-Lyttleton prescription \citep{HoyleLyttleton,BondiHoyle}, since in these cases the wind velocity is much larger than the orbital velocity of the accretor.
On the other hand, winds from AGB stars can have velocities smaller or comparable to the orbital velocity of the accretor.
In such situations, hydro-dynamical simulations have shown that efficient wind accretion is possible through the wind Roche lobe overflow (wRLOF) mechanism \citep[][]{Mohamed2007}, which is a regime between the Bondi-Hoyle-Lyttleton regime and RLOF.
In other words, in the wRLOF mechanism it is assumed that the slow winds fill the donors Roche lobe, which implies that wind accretion is enhanced.

%
Given the importance of this mechanism on the binary evolution we investigate here, we implemented the wRLOF model in \mesa, as described in \citet{Abate2013}, \citet{Ilkiewicz2019} and \citet{Belloni2020}.
Regarding how much of the donor wind is allowed to be accreted, since the accretor cannot accrete more mass than is lost by the red giant star, we enforced that the accretion rate efficiency in the Bondi-Hoyle-Lyttleton regime cannot be greater than $80$~per cent.
In addition, we assume that the accretion rate efficiency in the wRLOF regime has to be $\leq50$~per~cent, which is consistent with hydro-dynamical simulations \citep[see][and references therein]{Abate2013}.
Finally, during each step of the simulations, after calculating the accretion rate efficiency in both regimes, we adopted the higher one to be used in the wind mass transfer scheme in \mesa.

%
The zero-age main sequence (ZAMS) stars are assumed to have no rotation.
However, as the stars evolve, their rotations are allowed to change according to \mesa~standard prescriptions \citep[][]{Paxton2013}.
In addition to rotation, we also allow the stars to eventually synchronize with the orbit, due to tidal interaction.
As usual, we distinguish the phases in which the star has a radiative envelope from those in which it has a convective envelope since the synchronization time-scales in the former case are generally orders of magnitude larger than those in the latter.
In particular, given the star masses we investigate, we assume that main sequence and core-helium-burning stars have radiative envelopes, while sub-giant, FGB and AGB stars have convective envelopes.
During each star evolution, the synchronization time-scales are computed using the \mesa~standard prescription, which are based on \citet[][]{Hut1981} and \citet[][]{hurleyetal02-1}.

%
When the initially more massive star becomes a white dwarf, we ignore the contribution of the white dwarf to the tidal interaction, as its synchronization time-scale should be much longer than that of the initially less massive star.
However, since the white dwarf can accrete mass and angular momentum from the winds of its companion, especially 
in the case of a symbiotic binary with the giant donor being on the AGB, we allowed the WD to spin up following \citet{kingetal91-1} and \citet{schreiberetal21-1}.
Measuring short-time variability in symbiotic stars is very difficult, due to the fact that steady nuclear burning on the white dwarf surface usually reduces the amplitude of the variation, reducing in turn its detection probability.
Despite that, there is some evidence that the white dwarf can spin up during the symbiotic star phase from the studies of the stable periodic oscillations in two systems. 
In BF~Cyg, the white dwarf's spin period is ${\approx108}$~min \citep[][]{Formiggini2009}, while Z~And hosts a magnetic white dwarf with spin period of ${\approx28}$~min \citep[][]{Sokoloski1999}.
In both cases, the white dwarf most likely spun up due to wind mass and angular momentum accretion.

%
Depending on the accretion rate onto the white dwarf, hydrogen shell burning could be stable, which results in an increase of its mass \citep[e.g.][]{Shen2007,Wolf2013}.
We implemented the critical accretion rate derived by \citet[][their eq.~5]{Nomoto2007}, above which white dwarfs are thermally stable, i.e. hydrogen burns steadily in a shell.
For accretion rates lower than this critical value, the white dwarf is considered thermally unstable, which means that the hydrogen shell burning is unstable to flashes and, for sufficiently strong flashes, nova eruptions are triggered, causing that most of the accreted mass will be expelled. 
In addition, we assume that there is a maximum possible accretion rate \citep[][their eq.~6]{Nomoto2007} such that white dwarfs accreting at rates above it will burn stably at this maximum rate, and the remaining non-accreted matter will be piled up to form a red-giant-like envelope.
Since we expect such high rates when the TP-AGB companion of the first white dwarf is close to filling its Roche lobe, i.e. the binary is close to the onset of common envelope evolution, we assume that this piled-up matter will be eventually lost during common envelope evolution. 

Finally, we assume solar metallicity and set all other parameters in \mesa~as the default values in version 15140.

\subsection{Searching for a model that can explain the properties of NLTT~12758}

\begin{figure*}
\includegraphics[width=0.45\textwidth]{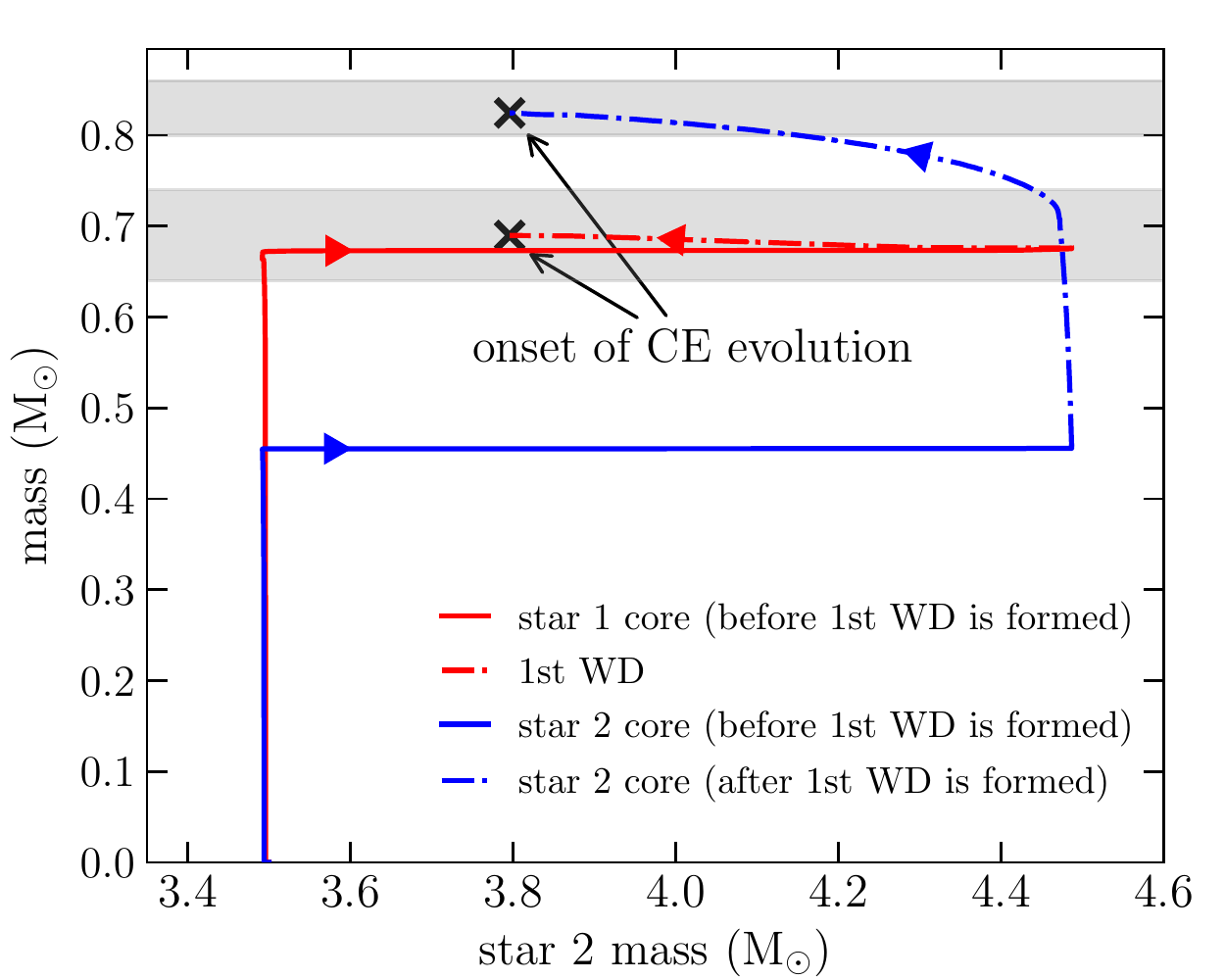}
\includegraphics[width=0.45\textwidth]{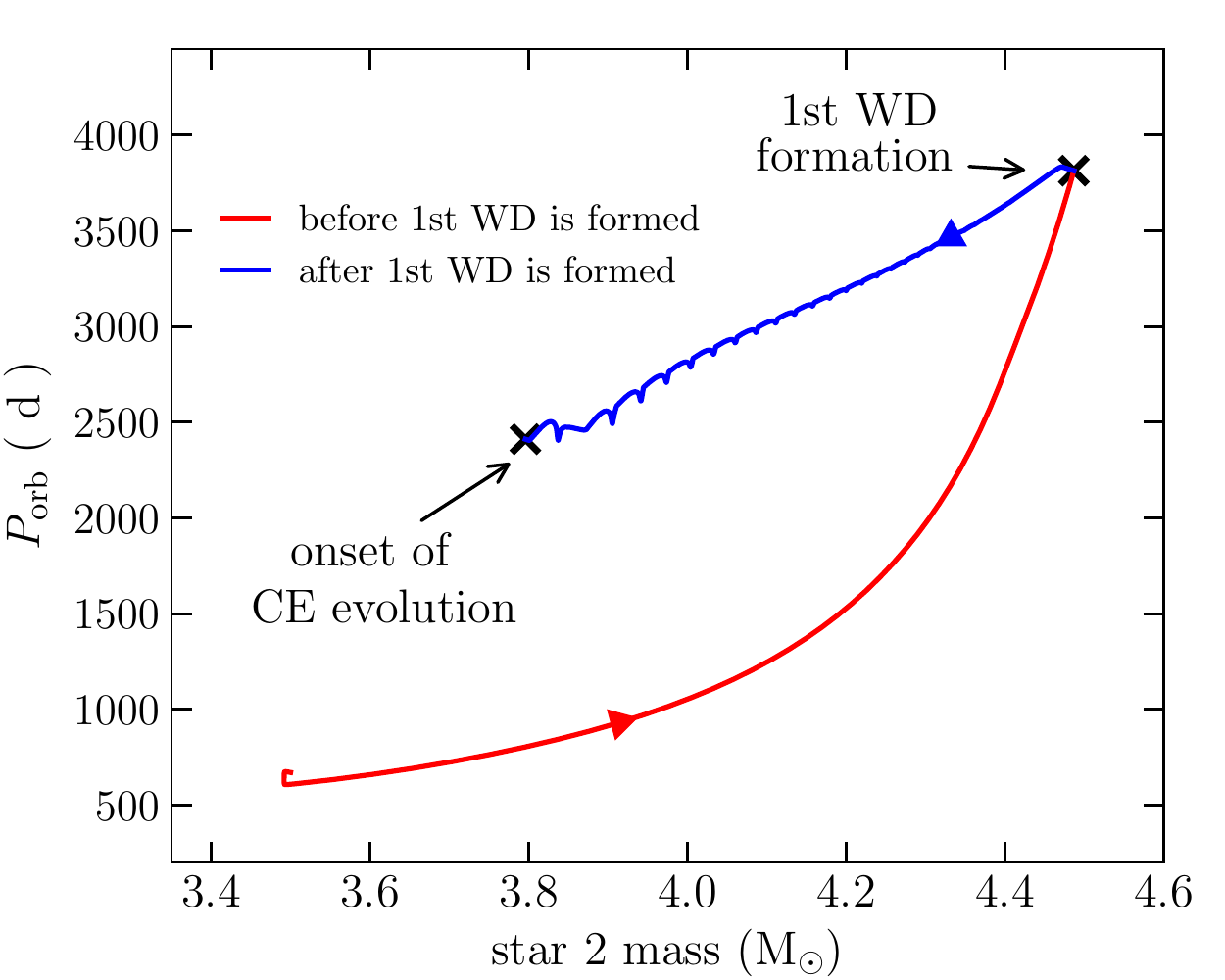}
\includegraphics[width=0.45\textwidth]{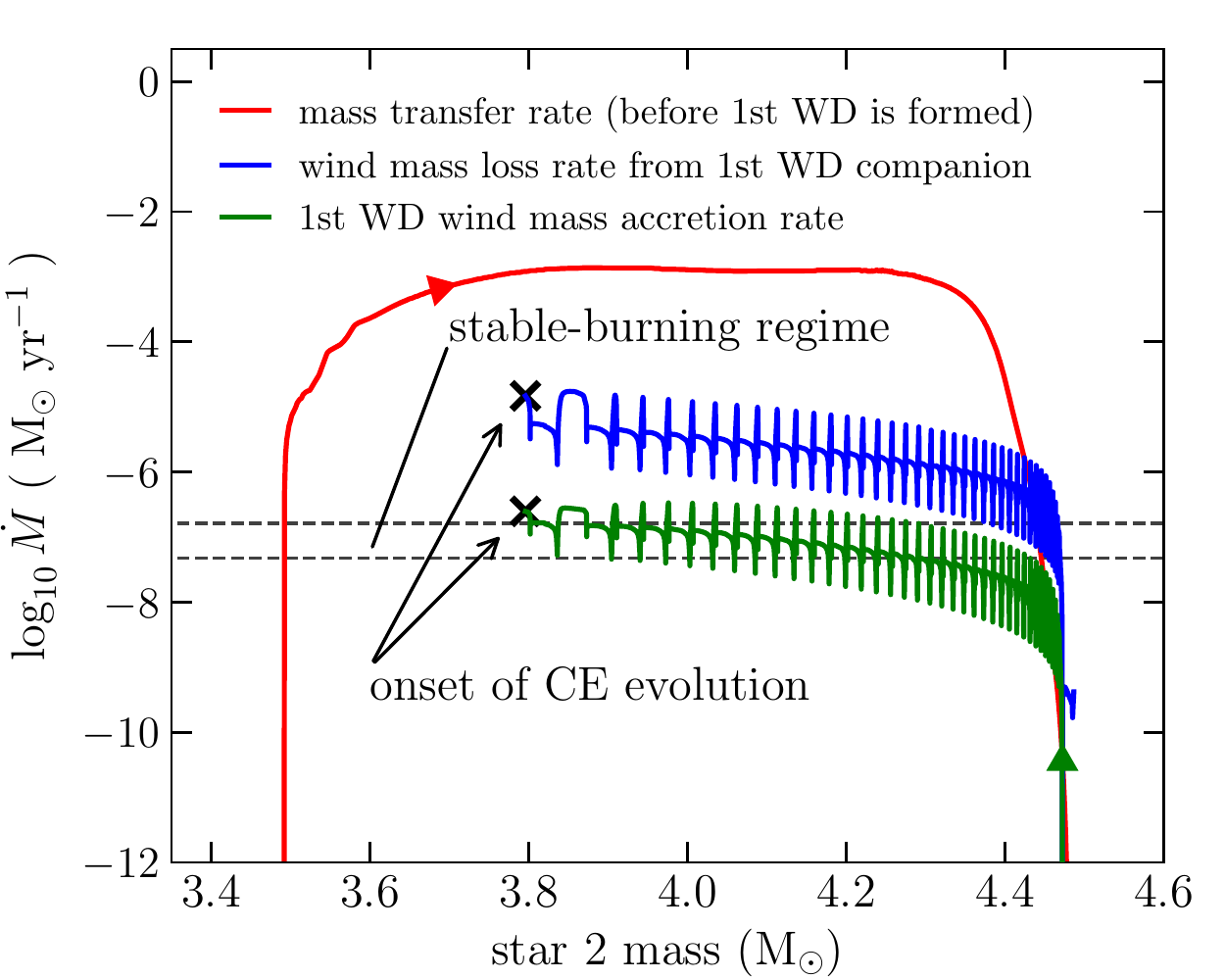}
\includegraphics[width=0.45\textwidth]{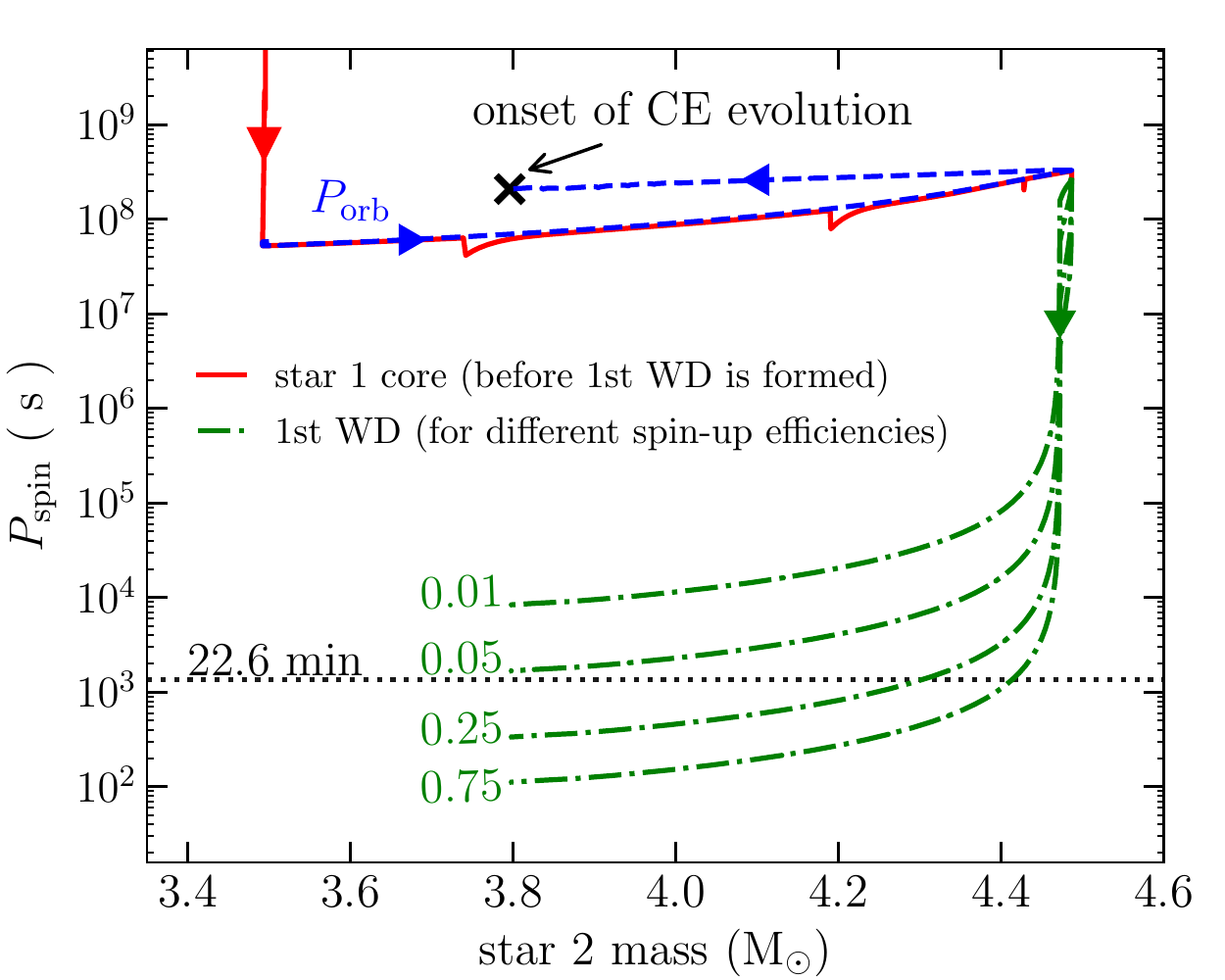}
\caption{Evolution with the DA progenitor (star 2) mass of some properties of the best-fitting model for NLTT\,12758, prior the onset of the common envelope phase, namely star core and 1st WD masses (top left-hand panel), orbital period (top right-hand panel),  mass transfer rate during stable non-conservative mass transfer and wind mass loss/accretion after the 1st WD is formed (bottom left-hand panel), and spin period of the DAP (1st WD) progenitor as well as of the DAP WD (bottom right-hand panel). The range of DA and DAP WD masses, as inferred from observations, are indicated as gray areas in the top left-hand panel, and the spin period of the DAP WD as a dotted horizontal line in the bottom right-hand panel. We reasonably well reproduce those properties in our modeling with the \mesa~code.}
\label{fig:MESA:evolution}
\end{figure*}

%
In order to search for a reasonable model that could reproduce the observational properties of NLTT\,12758 we carried out several sparse grids covering different regions of the parameter space, which include the ZAMS masses, the initial orbital period, the accretion efficiency during stable non-conservative mass transfer, and the orbital energy conversion efficiency during common envelope evolution ($\alpha_{\rm{CE}}$).

%
We found out that, if the magnetic WD in NLTT\,12758 was formed through stable non-conservative mass transfer, then its progenitor ZAMS mass cannot be less massive than ${\simeq3.60}$~M$_\odot$. 
This is because the core mass growth during the first white dwarf progenitor evolution is virtually entirely interrupted when it becomes a Roche-lobe filling AGB star.
This implies that the 1st white dwarf progenitor core mass at the onset of RLOF must be already comparable to the observed value of the DAP white dwarf in NLTT\,12758.
This also implies that the 1st white dwarf progenitor ZAMS mass cannot be much larger than ${\sim4.00}$~M$_\odot$, since stars with masses larger than that would develop a more massive core before the onset of RLOF.

In addition, stable mass transfer from an evolved red giant only occurs when the ZAMS star masses are comparable, which means that the mass ratio (more massive over the less massive) has to be $\lesssim1.1$.
Otherwise, the mass transfer will be most likely dynamically unstable.
Moreover, to reproduce the magnetic white dwarf mass and to have stable mass transfer, the red giant has to be on the early AGB. 
In case the mass transfer starts when the magnetic white dwarf progenitor is on the thermally pulsing AGB phase, the mass transfer will be most likely dynamically unstable, given the huge changes in the star size during this phase.

%
Regarding the initial orbital period, assuming that the onset of RLOF occurs when the star is on the early AGB phase implies that the initial orbital period cannot be longer than ${\sim700}$~d.
On the other hand, in order to reproduce the non-magnetic white dwarf mass in NLTT\,12758, the core mass of its progenitor needs to substantially grow before the onset of common envelope evolution.
This implies that when stable non-conservative mass transfer ends the orbital period has to be ${\sim3500-4000}$~d, which in turn implies that the initial orbital period of the ZAMS binary cannot be shorter than ${\sim500}$~d.
However, we shall emphasize that this lower limit for the initial orbital period strongly depends on the accretion efficiency during stable mass transfer.
In particular, the larger the accretion efficiency, the longer the orbital period at the end of stable mass transfer.
This correlation comes from the fact that the increase in orbital period due to mass transfer is much stronger than the increase caused by mass loss from the system.

%
To have dynamically stable mass transfer when the donor is an early AGB star, we found that the accretion efficiency must be smaller than ${\sim0.5}$, which means that at least half of the mass leaving the early AGB donor is not accreted by the donor, i.e. is lost from the binary.
Such huge mass loss may seem unrealistic at first glance, but there are several mechanisms able to drive mass loss from double red giant binaries, such as jets, stellar winds, circumbinary disc, accretion disc flashes/outbursts/outflows. Recently, while modeling the system 2M17091769$+$312758, which is a semi-detached binary composed of a red giant star transferring mass to a sub-giant star, \citet{milleretal21-1} found that to reproduce the observational properties of this system, around half of the mass leaving the donor star actually escapes the binary, corresponding to an accretion efficiency of $\sim0.5$ which is consistent with our finding.
However, we of course agree with one of the greats in the field who stated concerning mass loss that "{\em{details
of this process remain obscure}}" \citep{webbink08-1}.
%

\subsection{Reproducing the evolutionary history of NLTT~12758}

%

\begin{figure}
\includegraphics[width=0.45\textwidth]{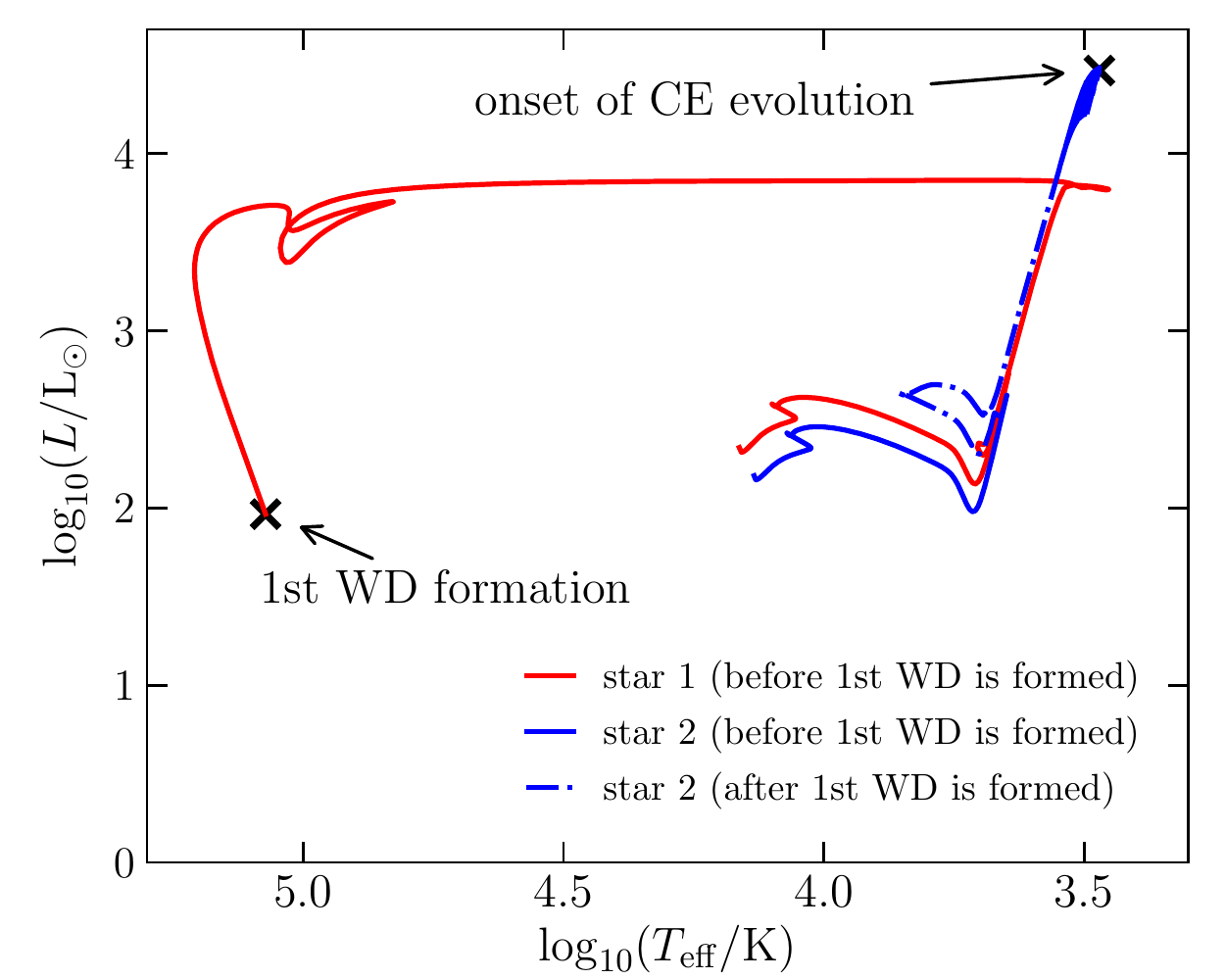}
\caption{HR diagram of both stars until the white dwarf formation. The initially lower mass star (blue) is already on the first giant branch when mass transfer starts (dashed dotted). At the onset of common envelope evolution, the initially lower mass star evolved higher up the AGB than the first star. This way, the higher mass white dwarf is formed second. }
\label{fig:MESA:HRD}
\end{figure}

%

The best-fitting model in our approach, combining dynamically stable non-conservative mass transfer followed by dynamically unstable mass transfer, is illustrated in Fig.~\ref{fig:NEW}. 
The ZAMS stars have masses of 3.85\,\Msun\, and 3.50\,\Msun\, and the orbital period is 670 days. The first mass transfer is stable but highly non-conservative. During this mass transfer phase the lower mass white dwarf of the system is formed, the secondary (the initially less massive) star reaches 4.49\,\Msun, and the period increases to 3814 days. 

As soon as this star evolves into a TP-AGB star, the binary becomes a symbiotic star and the white dwarf efficiently accretes mass and angular momentum from its wind and spins up, reaching the observed values of the mass (0.69\,\Msun) and spin period ($\sim$23 minutes). Due to tidal forces, which synchronize the rotation of the giant star with the orbital motion, the orbital period decreases during this symbiotic phase.
This finding is consistent with observational indications that most red giants in symbiotic stars are synchronized with the orbit \citep[][]{Zamanov2007}. 

As soon as the giant fills its Roche-lobe dynamically unstable mass transfer leads to common envelope evolution which then leaves behind a close double white dwarf with a fast spinning 0.69\,\Msun and a more massive 0.825\,\Msun component.  
For a common envelope efficiency of 0.3, considering no contributions from recombination energy, and using $\lambda$ as calculated by \bse, the orbital period of this emerging double white dwarf binary is $1.16$\,days. Within the following $\sim2$\,Gyr its period decreases to $1.154$\,days due to angular momentum loss by the emission of gravitational radiation. The age difference between the two white dwarfs is just 50 Myr. The two white 
dwarfs cool and start to crystallize, at first the more massive DA white dwarf which does not generate a strong magnetic field as this white dwarf is not rotating rapidly and roughly $\sim0.7$\,Gyr later the lower mass white dwarf. When the latter happens, the rotation and crystallization driven dynamo generates the strong magnetic field of the DAP white dwarf we observe today.  

The evolution of both stars prior to the final common envelope phase is further illustrated in Fig.~\ref{fig:MESA:evolution}. We plotted the core masses, orbital period, mass transfer rate, and the spin period as a function of the mass of the initially less massive star. 

As shown in the upper left panel, the mass of the white dwarf that later forms the magnetic component (the white dwarf that formed first) increases between the main two mass transfer events due to wind accretion (red dashed--dotted line). The core mass of the initially less massive star (blue line) has already reached more than $0.4$\,\Msun\, when stable mass transfer starts but increases further after the mass transfer phase ends, until the system enters common envelope evolution. 

The upper left plot illustrates how the orbital period changes during the evolution. During the stable mass transfer phase the period significantly increases (red line) but it decreases before the onset of common envelope evolution due to tidal forces when the initially lower mass star evolves on the AGB. 

The evolution of the mass transfer is shown in bottom left panel of Fig.~\ref{fig:MESA:evolution}. The red line shows that the stable mass transfer reaches large values of $\sim10^{-3}$\,\Msun/yr, and that roughly three per cent of the wind mass loss rate (blue) is accreted by the white dwarf (green) during the symbiotic phase. This mass accretion leads to the spin-up of the white dwarf that formed first as displayed in the bottom left panel. The final spin period depends therefore on the mass transfer rate and on the assumed spin-up efficiency. 

The full evolution of both stars prior to the white dwarf formation is also illustrated in an HR diagram (Fig.~\ref{fig:MESA:HRD}). The mass ratio of both stars is so close to one that the lower mass star is already on the 
FGB when the first white dwarf is formed. During the stable mass transfer phase, as the FGB accretor star is growing in mass, it also rejuvenates and becomes hotter.

The very good agreement in the stellar masses, orbital and spin period, and of the ages of the white dwarfs is of course the result of adjusting the values of the common envelope efficiency, the mass loss parameter, the initial masses and orbital period as well as the angular momentum accretion efficiency. However, we would like to emphasize that all other parameters for stellar evolution and mass transfer are fixed and that the presented solution does not represent a rare or unusual evolution. 
In fact, we found many solutions that are in very good agreement with the observations and could easily produce a broad population of double white dwarfs that formed through the channel we propose for NLTT\,12758. In addition, any model for the formation of 
double white dwarfs that we are aware of contains at least the same number of free parameters.

The agreement between theoretical predictions and observations of the DAP white dwarf in NLTT\,12758 is as good as it gets. Our new evolutionary scenario combined with the rotation and crystallization driven dynamo scenario explains the magnetic field generation in NLTT\,12758. This dynamo mechanism is also consistent with the observational result that most white dwarfs that are members of close double white dwarfs are not magnetic because the vast majority of them are not crystallizing.  
The only exceptions are the DA companion to the magnetic white dwarf in NLTT\,12758 and the more massive components in NLTT\,11748 and SDSS\,J1257+5428. 
In the next section, we briefly discuss these white dwarfs. 

\section{Should the massive components in NLTT~12758, NLTT~11748 and SDSS~J1257+5482 be magnetic?} 

The more massive DA white dwarf in NLTT\,12758 is clearly crystallizing but according to our evolutionary scenario has never had the possibility to accrete significant amounts of angular momentum. According to the 
rotation and crystallization driven dynamo scenario it is therefore not surprising that this white dwarf is not strongly magnetic. 

In the case of NLTT\,11748 the situation is more difficult to evaluate because neither the evolutionary history of the system nor the magnetic or non-magnetic nature of the massive white dwarf are known.  
Spectral observations of NLTT\,11748 by \citet{kawka+vennes09-1} revealed the brighter component to be a low-mass ($< 0.2$\Msun) helium 
core white dwarf and \citet{steinfadtetal10-1} demonstrated that the companion is a relatively massive C/O white dwarf.
\citet{kaplanetal14-1} analyzed high-precision eclipse light curves and found the 
low mass white dwarf to be hotter ($T_{\mathrm{eff}}\simeq8700\pm140$\,K) and with an H-rich (DA) atmosphere while the companion is cooler 
($T_{\mathrm{eff}}=7600\pm120$\,K) and could therefore be crystallizing. 
Unfortunately, the more massive and cooler white dwarf is not visible in the 
spectra that have been taken of this object and it is therefore impossible to tell whether it hosts a strong magnetic field or not. 

It also remains unclear which of the two white dwarfs formed first. White dwarfs with masses below $0.2$\,\Msun, such as the brighter component in 
NLTT\,11748, can experience stable H burning for 
several Gyr \citep{driebeetal99-1,serenellietal02-1,paneietal07-1}. We can therefore not fully exclude that the cooler and more massive white dwarf formed after the low-mass white dwarf. This scenario requires the first mass transfer to be stable, similar to the scenario we developed in this paper which was inspired by earlier work \citep[e.g][]{woodsetal12-1}. In this case, the more massive white dwarf might not have accreted sufficient mass and angular momentum and the dynamo scenario would not apply. If, in contrast, the more massive white dwarf formed first, the accretion of some mass and angular momentum during the second mass transfer phase could be expected and the conditions for the dynamo to work would be met.

The third double white dwarf with one component most likely crystallizing is SDSS\,J1257+5482. In this system the lower mass white dwarf seems to be significantly older than the more massive component. It is currently not clear how this system formed \citep{boursetal15-1} and we therefore restrain ourselves from a discussion of the potential accretion of mass and angular momentum of the massive component of the system.  

Interestingly, both \citet{kulkarnietal10-1} and \citet{marshetal11-1} found that the Balmer lines of the massive component 
are broadened significantly relative to what is expected from just pressure broadening. Mechanisms that have been suggested for the additional broadening are fast rotation and/or a magnetic field. It might thus be that the massive component in SDSS\,J1257+5428 is the second fast spinning and crystallizing magnetic white dwarf in a double degenerate binary. Testing this hypothesis would provide important constraints on the evolutionary history of the binary. 

Given the above discussion, we conclude that the predictions of the crystallization and rotation driven dynamo scenario are consistent with the observed incidence of magnetism among close double white dwarfs. 
This separates the dynamo from the common envelope dynamo scenario previously suggested. According to the common envelope idea for the origin of the magnetic fields in white dwarfs 
\citep[e.g.][]{toutetal08-1}, a large number of hotter white dwarfs 
that are members of double white dwarfs should be magnetic as shown by \citet{belloni+schreiber20-1}. This does not seem to be the case: among the relatively well studied 
57 double white dwarfs listed in Table\,\ref{tab:dwds} only for one white dwarf, the fast spinning component in NLTT\,12758, the detection of a magnetic field has been reported.

\section{Could all (strong) magnetic fields in white dwarfs be generated by the dynamo?} 

While representing a promising scenario for a relatively large number of magnetic fields in white dwarfs in different settings that are otherwise inexplicable, the rotation and crystallization driven dynamo is certainly not the only mechanisms producing strongly magnetic white dwarfs. As illustrated in figure\,16 of \citet{ferrarioetal15-1}, white dwarfs with field strength exceeding 100\,MG are found to be slow rotators which is inconsistent with one of the conditions proposed for the dynamo. In addition, carefully inspecting the full sample of known magnetic white dwarfs in close binaries reveals that not even all magnetic fields of white dwarfs in these systems can be explained by the dynamo. 

The magnetic CV HY\,Eridiani contains a white dwarf that is most likely consisting of a He-core as its mass is below $0.47\Msun$ at a 90 per cent confidence level. The dynamo proposed by \citet{schreiberetal21-1}, however, relies on the crystallization of a C/O core which appears unlikely (although not impossible) for HY\,Eridiani.

One of the biggest strengths of the crystallization and rotation driven dynamo scenario is that is explains the large number of magnetic CVs and the absence of strongly magnetic white dwarfs among young detached post common envelope binaries. However, in two cases, weakly magnetic white dwarfs have been detected in young close detached white dwarf binaries with main sequence star companions. 

The first example is the V471\,Tau, with a estimated field strength of a few hundred kG \citep{sionetal12-1}, 
and a white dwarf temperature of $34500$\,K which clearly excludes crystallization and thus the dynamo to be operating. An alternative scenario that can be excluded to be the dominant formation mechanism of magnetic white dwarfs might be at work here: fossil fields. V471\,Tau is most likely the descendant of a compact triple star system as the white dwarf is the youngest and most massive white dwarf in the Hyades \citep{obrienetal01-1}. This potential history offers an explanation for the magnetic nature of the white dwarf as magnetic main sequence Ap/Bp stars are likely the result of main sequence star mergers \citep{ferrarioetal09-1,schneideretal19-1}. It might therefore be that the origin of the magnetic field of the white dwarf in V471\,Tau is best explained by the fossil field scenario. 

The second magnetic white dwarf in a young close detached binary, CC\,Ceti, was discovered only recently with a field strength of $600-700$kG \citep{wilsonetal21-1}. The white dwarf in this system is most likely a He-core white dwarf and even if it was not, it is clearly too hot ($25203\pm42$,K) to be crystallizing.  

While the crystallization and rotation driven dynamo offers a consistent explanation for 
the origin of the magnetic fields in many white dwarfs, the examples listed above clearly demonstrate that it cannot be the only mechanism producing magnetic white dwarfs. For massive and young magnetic single white dwarfs stellar or common envelope merger events may offer an explanation. The magnetic field in V471\,Tau might be of the fossil field origin as the progenitor star of the white dwarf was likely formed by a stellar merger. 
The origin of the magnetic fields in the close binaries CC\,Ceti and HY\,Eri remains unclear.

\section{Concluding discussion}

The magnetic white dwarf in NLTT\,12758 is rapidly rotating 
and its temperature and mass are consistent with the core having started to crystallize. 
Therefore, the crystallization and rotation driven dynamo can explain the occurrence of the detected magnetic field. 
Furthermore, the model also explains that in all other known double white dwarfs no magnetic component has been found.

To understand the fast rotation of the magnetic white dwarf in NLTT\,12758, we investigated the evolutionary history of the system and found that a previously suggested sequence consisting of two common envelope events appears to be unlikely as additional energy sources are required and given that the cooling ages of both white dwarfs are rather similar. 

We propose that instead a phase of stable mass transfer, a symbiotic phase during which the first formed white dwarf accretes mass and angular momentum, and common envelope evolution as the second main mass transfer phase led to the formation of NLTT\,12758. We performed MESA simulations that illustrate how well this revised evolutionary scenario can reproduce the observations of the system. 
Our simulations show how the combination of stable mass transfer and a subsequent common envelope event can lead to a double white dwarf binary consisting of two relatively massive C/O white dwarfs.  
In particular, and most importantly in the context of this paper, they offer an explanation for the increased rotation rate of the magnetic white dwarf, which represents a key ingredient for magnetic field generation according to the rotation and crystallization driven dynamo. 

Our finding that the dynamo scenario can consistently explain the magnetic nature of NLTT\,12758 (and the absence of clear signs of magnetism among close double white dwarfs otherwise) adds another piece of evidence to the already available support for the dynamo scenario for the generation of strong magnetic fields in white dwarfs: the dynamo is the only scenario that can explain the absence of strongly magnetic white dwarfs among young post common envelope binaries, it offers an explanation for the existence of the radio pulsing white dwarf binary AR\,Sco, it is consistent with the high occurrence rate of magnetic white dwarfs in cataclysmic variables \citep{schreiberetal21-1}, it naturally explains the absence of high accretion rate intermediate polars in globular clusters \citep{bellonietal21-1}, 
and seems to be the reason behind the observed relation between magnetism and metal pollution \citep{schreiberetal21-2}. 
Considering that also single white dwarfs that are currently not metal polluted, might have accreted planetary material in the past and therefore gained angular momentum \citep{schreiberetal21-2}, the model might also explain the 
magnetism of old white dwarfs that currently do not show signs of accretion \citep{bagnulo+landstreet21-1}. 

All this recent evidence shows that the idea initially put forward by \citet{isernetal17-1} and further developed by \citet{schreiberetal21-1}
has the potential to significantly help to finally solve the long standing mystery of the origin of strong magnetic fields in white dwarfs.  

However, despite this success, the case cannot be closed yet. 
Magnetic white dwarfs exist that the dynamo model fails to explain. This concerns slowly rotating magnetic single white dwarfs as well as three magnetic white dwarfs in close binary stars (V471\,Tau, CC\,Ceti, HY\,Eri) which either contain He-core white dwarfs 
or a non-crystallizing C/O white dwarf. At least for these objects, alternative scenarios
for the generation of magnetic fields in white dwarfs need to be considered.

In addition, on the modeling side, the scenario remains rather phenomenological as we 
currently do not have theories that allow us to 
properly calculate how long it takes the magnetic 
field to emerge as soon as the dynamo kicks in. We also have no clue if there are upper and lower limits for the crystallized mass fractions for the dynamo to properly work (and if so, what are their values?). 
Finally, despite the rather detailed discussion presented in \citet{schreiberetal21-1}, proper population synthesis that include magnetic field generation through the dynamo have still to be performed. 

From an observational point of view, we need to expand the work of \citet{bagnulo+landstreet21-1} who presented the first volume limited (but still rather small) 
sample of white dwarfs and investigated how the incidence of magnetism depends on key parameters. 
As in SDSS-V white dwarfs are for the first time prime targets of 
an SDSS survey, we may indeed be able to analyze a large volume limited sample of white dwarfs in the near future. Combining this sample with population models of single white dwarfs and those that are part of binary systems would provide key constraints on the dynamo scenario and should bring us closer to finally understand the origin of strongly magnetic white dwarfs.

\section*{Acknowledgements}
We thank the researchers that provided the microphysics necessary to develop \mesa. The \mesa~equation of state is a blend of the OPAL \citep{Rogers2002}, SCVH \citep{Saumon1995}, PTEH \citep{Pols1995}, HELM \citep{Timmes2000}, and PC \citep{Potekhin2010} equation of states.
Radiative opacities are primarily from OPAL \citep{Iglesias1993,Iglesias1996}, with low-temperature data from \citet{Ferguson2005} and the high-temperature, Compton-scattering dominated regime by \citet{Buchler1976}.
Electron conduction opacities are from \citet{Cassisi2007}. Nuclear reaction rates are a combination of rates from NACRE \citep{Angulo1999}, JINA REACLIB \citep{Cyburt2010}, plus additional tabulated weak reaction rates \citep{Fuller1985, Oda1994,Langanke2000}.
Screening is included via the prescription of \citet{Chugunov2007}. Thermal neutrino loss rates are from \citet{Itoh1996}. 

MRS and MZ acknowledge support from Fondecyt (grant 1221059).  
MRS is furthermore spported by ANID, -- Millennium Science Initiative Program -- NCN19\_171.
DB was supported by ESO/Gobierno de Chile. BTG was
supported by the UK STFC grant ST/T000406/1. 
SGP acknowledges the support of a STFC Ernest Rutherford Fellowship.
This project has received funding from the European Research Council (ERC) under the European Union’s Horizon 2020 research and innovation programme (Grant agreement No. 101020057). 

For the purpose of open access, the author has applied a Creative Commons Attribution (CC BY) licence to any Author Accepted Manuscript version arising.

\section*{Data Availability}

The MESA files required to reproduce our simulations will be provided upon request.






\appendix
\section{Close double white dwarf binaries} 

We compiled a comprehensive list of close (periods below 35 days) 
double white dwarfs. The resulting Table\,\ref{tab:dwds} should be complete with respect to double white dwarfs that contain a C/O white dwarf component with measured effective temperature.
Such systems are highlighted in bold and represent those that are plotted in Figure\,1 of 
this paper. 

We did not include all systems from the recently published large sample 
of almost hundred extremely low-mass (ELM) white dwarfs \citep{brownwetal20-1} as in all but one of these systems the stellar components with a measured temperature are helium core low-mass white dwarfs which are irrelevant for the rotation and crystallization driven dynamo. 
The only exception, SDSS\,J1638+3500, is included in the table.

\small
\begin{table*}
    \centering
    \caption{Parameters of the known close double white dwarfs with measured orbital periods. We ignored many ELM systems where the only component with measured temperature is certainly a helium core white dwarf. For a comprehensive list of such systems see \citet{brownwetal20-1}. Components highlighted in bold are (potential) C/O white dwarfs with measured $T_{\mathrm{eff}}$ which are the white dwarfs of relevance for this paper and which are plotted in Fig.\ref{fig:M-Teff}. }
	\label{tab:dwds}
	\renewcommand{\arraystretch}{0.96} 
    \begin{tabular}{lcccccr}
        \hline
         Name & $M_1[\mathrm{M_{\odot}}]$ & $M_2[\mathrm{M_{\odot}}]$ & $T_{\mathrm{eff},1}[K]$ & $T_{\mathrm{eff},2}[K]$ & Period & Reference \\
        \hline
        NLTT11748   (WD 0342+176)  &  
        0.136-0162 & {\bf 0.707-0.740} & 8706$\pm$136 & {\bf 7597$\pm$119}$^\dagger$ & 0.24\,day & 1\\
        CSS 41177 & 0.38$\pm$0.02 & 0.32$\pm$0.01 & 24500  &  11500 &  2.78\,hr & 2,3 \\ 
        GALEXJ171708.5+675712 & 0.185$\pm$0.01 & $\geq$0.86 & 14\,900$\pm$200  & -- &  5.90\,hr & 4\\
        SDSS\,J065133.33+284423.3 & 0.25 &  0.55 &  16\,400$\pm$300 & -- & 12.75\,min. & 5 \\
        SDSS\,J075141.18-014120.9   & 0.19$\pm$0.02 & 0.97$^{+0.06}_{-0.01}$ & 15\,750$\pm$240 & -- &  1.92\,hr & 6 \\  
        SDSS J115219.99+024814.4   & {\bf 0.47$\pm$0.11} & {\bf 0.44$\pm$0.09} & {\bf 25\,500$\pm$1000} & {\bf 14\,350$\pm$500} & 2.40\,hr & 7, 8 \\
        ZTF J153932.16+502738.8  & {\bf{0.61$^{+0.017}_{-0.022}$}} & 0.210$^{+0.014}_{-0.015}$ &  {\bf 48\,900$\pm$900} &  $\leq$10\,000 & 6.91\,min. & 9 \\
        ZTF J190125.42+530929.5  & -- & -- & 28\,000$\pm$500 & 17\,600$\pm$400 & 40.60\,min. & 10 \\ 
        ZTF J2243+5242  &  0.349$^{+0.093}_{-0.074}$  &   0.384$^{+0.114}_{-0.074}$ &  22\,200$^{+1800}_{-1600}$ &  16\,200$^{+1200}_{-1000}$ &  8.8\,min & 11\\
        PG 1632+177 & 0.392$^{+0.069}_{-0.059}$ &  {\bf 0.526$^{+0.095}_{-0.082}$} &  8\,800$\pm$500 &  {\bf 11\,200$\pm$500} &  2.05\,day & 12\\
        WD 1534+503  & 0.392$^{+0.069}_{-0.059}$ & {\bf 0.617$^{+0.110}_{-0.096}$}  & 8\,900$\pm$500 &  {\bf 8\,500$\pm$500} &  0.71\,day & 12 \\
        SDSS J033816.16-813929.9 & 0.23$\pm$0.015 & 0.38+0.05-0.03 & 18\,100$\pm$300 & 10\,000$\pm$1000 & 30.6\,min. & 13 \\
        SDSS J063449.92+380352.2 &  {\bf 0.452$^{+0.070}_{-0.062}$} & 0.209$^{+0.034}_{-0.021}$ & {\bf 27\,300$^{+4000}_{-2900}$} & 10\,500$^{+300}_{-200}$ & 26.5\,min & 13 \\
        WD 1434+503  (SDSS J143633.29+501026.8)	&	0.23(01) & -- & 17\,120(200) & --	& 1.15\,hr & 14 \\
        WD 1050+522 (SDSS J105353.89+520031.0)  &  0.22(01) & --  & 16\,150(200) &--  &  0.96\,hr & 14 \\ 
        SDSSJ1337+3952 & {\bf 0.46$\pm$0.02} &  0.26$\pm$0.01  &  {\bf 9\,450$\pm$80}  &   7\,520$\pm$170 &  99\,min.  & 15\\  
        WD0028-474  & {\bf 0.60$\pm$0.06} &  0.45$\pm$0.04 & {\bf 18\,500$\pm$500} & 17\,000$\pm$500  & 9.35\,hr & 16\\ 
        SDSSJ0318–0107 &. 0.4$\pm$0.05 &  {\bf 0.49$\pm$0.05} & 14\,500$\pm$500 & {\bf 13\,500$\pm$500} & 45.9\,hr & 16\\
        HE0410-1137 & {\bf 0.51$\pm$0.04} & 0.39$\pm$0.03  &  {\bf 16\,000$\pm$500}  &   19\,000$\pm$500 & 12.2\,hr & 16\\
        {\bf{NLTT12758}} & {\bf 0.83$\pm$0.03} & {\bf 0.69$\pm$0.05} & {\bf 7950$\pm$50} & {\bf 7220$\pm$180} & 1.154\,day & 17\\
        WD1202+608 (Feige 55) & {\bf 0.3-0.487} &   $\geq$0.25 &   {\bf 56\,300$\pm$1000} &   -- &  1.493\,day  & 18 \\
        SDSSJ125733.63+542850.5 &  0.24  &   {\bf 1.06$\pm$0.05}   &  6400$\pm$50  &  {\bf 13\,030$\pm$150} & 4.6\,hr & 19 \\
        WD\,1704+481 & 0.39$\pm$0.05 & {\bf 0.56$\pm$0.07} & 9000 & {\bf 10\,000} & 0.145\,day & 20, 21\\
        WD0136+768 & {\bf 0.47} & 0.37 & {\bf 18\,500} & 10\,500 &1.41\,day & 21, 22\\
        WD1704+481 & 0.39$\pm$ 0.05 & {\bf 0.54} & 9000 & {\bf 10\,000} & 0.145\,day & 21\\    
        WD0957-666 & 0.37$\pm$0.02 & 0.32$\pm$0.03 & 30000  &  11\,000 &  1.46\,hr & 21, 23, 24 \\
        WD1204+450 & 0.46 & {\bf 0.52} & 31\,000 & {\bf 16\,000} & 1.603\,day & 21 \\
        WD0135-052 (L870-2) &  {\bf 0.47$\pm$0.05} & {\bf 0.52$\pm$0.05}  & {\bf 7470$\pm$500} & {\bf 6920$\pm$500} & 1.556\,day & 25, 26 \\
        WD1101+364  (=PG1101+364)  &  0.29  & 0.33  & 15\,500 & 12\,000 &  0.145\,day & 27 \\
        PG1115+166 & 0.70   &   0.70    &  -- & -- &  722.2\,hr & 28 \\ 
        SDSS J174140.49+652638.7 & 0.17$\pm$0.02 & $\geq1.11$ & 10540$\pm$170 & -- & 1.47\,hr & 6 \\
        WD0225-192 (HE0225-1912)  &  {\bf 0.55}   &   0.23 & {\bf 20488}  & -- &  0.22\,day & 29\\
        WD0315-013 (HE0315-0118)  & {\bf 0.50} & 0.49 & {\bf 12720} &  -- &  1.91\,day & 29\\ 
        WD0320-192 (HE0320-1917)  & 0.31 & 0.45 & 13\,248 & --  &   0.86\,day & 29, 30, 31\\ 
        WD0326-273  &  0.364 & $\geq$0.96 &  9158 & -- &  1.88\,day & 29, 30, 31 \\
        WD0453-295  & 0.40 & 0.44 & 16\,360 & 1330 & 0.36\,day & 29, 32\\
        WD1013-010  & 0.32  & $\geq$ 0.62 & 8080 & -- &  0.44\,day & 29, 30, 31\\ 
        WD1022+050  & 0.37  &  $\geq$0.28 & 14\,693 & -- &  1.16\,day & 29, 31, 33, 34\\  
        HS1102+0934 & 0.38  & $\geq$0.45 &  16\,961 & -- & 0.55\,day & 29, 30, 31, 35\\
        WD1210+140  & 0.33  &  $\geq$0.44 &  32\,127 & -- &  0.64\,day & 29, 30, 31\\  
        HS1334+0701 & 0.35  &  -- &  16\,891 & -- &  0.23\,day (uncertain) & 29 \\  
        WD1349+144  & {\bf 0.55} & 0.33  & {\bf 19917} & --&  2.21\,day & 29, 30, 31 \\ 
        HE1414-0848 & {\bf 0.52} &  0.74 & {\bf 11133} & -- & 0.52\,day & 29, 31, 34 \\  
        HE1511-0448 & {\bf 0.50} &  $\geq$0.67 & {\bf 50\,899} & -- & 3.22\,day & 29, 30, 31 \\ 
        WD1824+040  & 0.4 & $\geq$0.73 &  14\,787 & -- &  6.27\,day & 25, 29, 31, 34\\
        WD2020-425  & {\bf 0.81} & 0.54 & {\bf 34\,004} & -- &  0.3\,day & 29, 30 \\
        HE2209-1444 & 0.43 & 0.72 & 8471 & -- & 0.28\,day & 29, 31, 34\\
        WD1428+373 & 0.35 & $\geq$0.23 & -- & -- & 1.14\,day & 34 \\
        WD2032+188 & 0.41 & $\geq$0.47 & -- & -- & 5.08\,day & 34 \\
        SDSSJ0755+4800  & 0.42 &  $\geq$0.90 & 19\,890$\pm$350 & -- &  0.55\,day &  35 \\
        SDSSJ1104+0918  & 0.46 & $\geq$0.55 & 16\,710$\pm$250 & -- &  0.55\,day & 35 \\
        SDSSJ1557+2823  & {\bf 0.49} & $\geq$0.43 & {\bf 12\,550$\pm$200} & -- &  0.41\,day & 35 \\ 
        WD2331+290 & 0.39 & $\geq$0.32 & -- & -- & 0.17\,day & 36 \\ 
        WD1713+332 & 0.35 & $\geq$0.18 & --& --& 1.12\,day & 36 \\
        WD1241-010 & 0.31 & $\geq$0.37 & -- & --& 3.35\,day & 36\\
        WD1317+453 & 0.33 & $\geq$0.42 & -- & -- & 4.87\,day & 36 \\
        SDSSJ1638+3500 & {\bf 0.698$\pm$0.030} & -- & {\bf 37250$\pm$570} & -- & 0.91\,day & 37  \\
         \hline
    \end{tabular}
    \noindent
    \small{
    $\dagger$ values correspond to assuming a thin envelope. 
    \noindent
    References: 
(1) 
\citet{kaplanetal14-1}, 
(2) \citet{parsonsetal11-1},
(3) \citet{boursetal14-1},
(4) \citet{vennesetal11-1},
(5) \citet{brownetal11-2},
(6) \citet{kilicetal14-1},
(7) \citet{hallakounetal16-1},
(8) \citet{parsonsetal20-1},
(9) \citet{burdgeetal19-1}, 
(10) \citet{coughlinetal20-1},
(11) \citet{burdgeetal20-1},
(12) \citet{kilicetal21-1}, 
(13) \citet{kilicetal21-2},
(14) \citet{mullallyetal09-1}, 
(15) \citet{vedantetal21-1},
(16) \citet{rebassaetal17-1},
(17) \citet{kawkaetal17-1},
(18) \citet{holbergetal95-1},
(19) \citet{boursetal15-1},
(20) \citet{maxtedetal00-1}
(21) \citet{maxtedetal02-1}
(22) \citet{bergeronetal92-1},
(23) \citet{moranetal97-1},
(24) \citet{bragagliaetal90-1}
(25) \citet{safferetal88-1},
(26) \citet{bergeronetal89-1},
(27) \citet{marsh95-1},
(28) \citet{maxtedetal02-2},
(29) \citet{napiwotzkietal20-1},
(30) \citet{koesteretal09-1},
(31) \citet{nelemansetal05-1},
(32) \citet{wesemaeletal94-1},
(33) \citet{maxted+marsh99-1},
(34) \citet{morales-ruedaetal05-1},
(35) \citet{brownwetal13-1},
(36) \citet{marsh95-2},
(37) \citet{brownwetal20-1}
}
\normalsize
\end{table*}

%
%


\normalsize

\section{Understanding the BSE input parameters}

\begin{figure*}
\includegraphics[width=0.7\textwidth,angle=270]{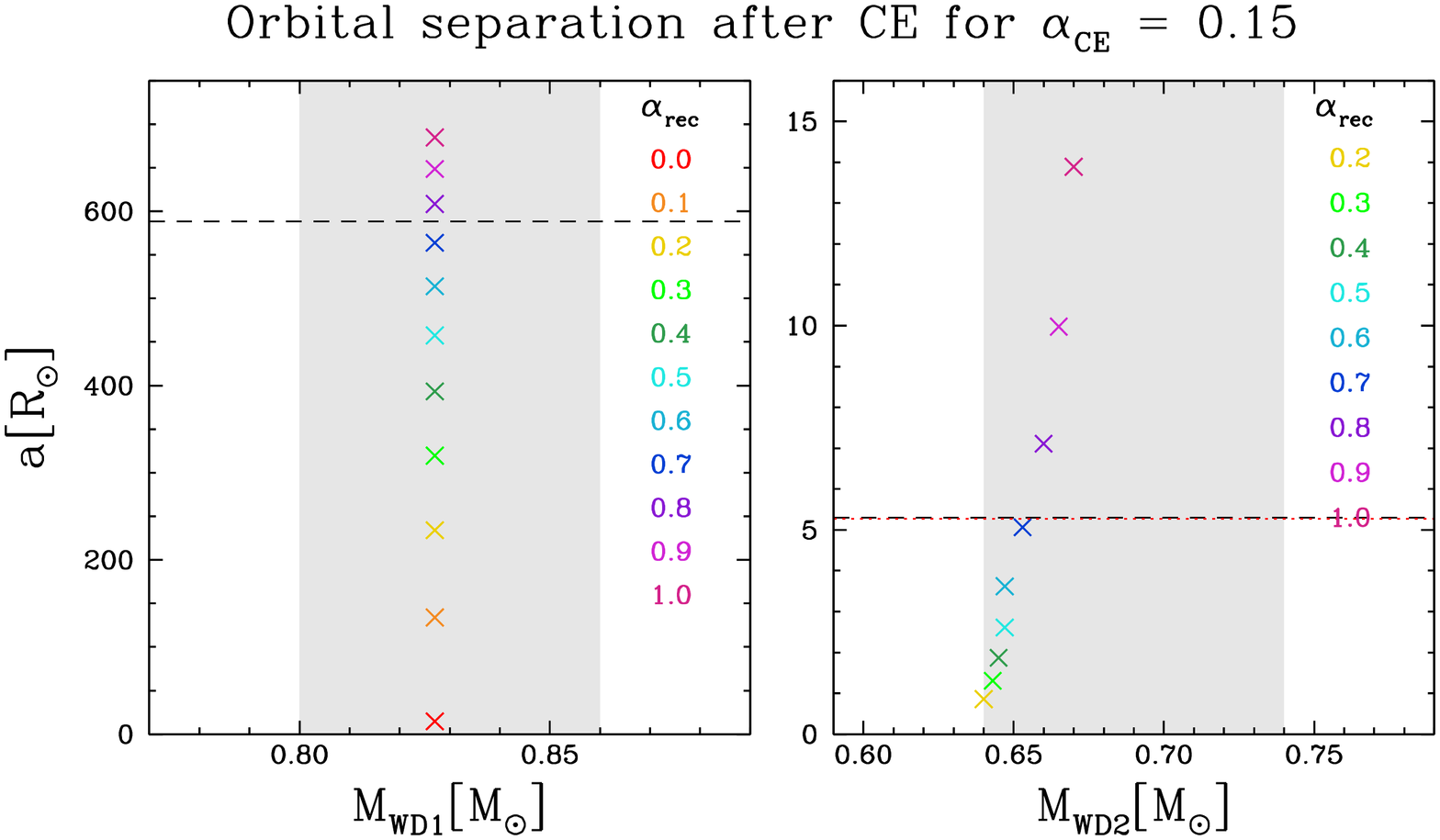}
\caption{White dwarf mass versus orbital separation after first (left) and second (right) common envelope phase for different fractions of the ionization/recombination energy. The gray area is for the measured white dwarf masses (with their error), the black dashed lines correspond to the orbital separations obtained by \citep{kawkaetal17-1} after each of the two common-envelope episodes, while the red dotted line in the right panel is the current orbital separation \citep{kawkaetal17-1}. Apparently (and potentially without being aware of it) between 70 and 80 per cent of the available recombination energy was assumed to contribute to expelling the envelope which can be considered as unrealistic.}
\label{fig:NLT015}
\end{figure*}

According to the \citet{kawkaetal17-1}, agreement with the observed characteristics of NLTT\,12758 was achieved with {\bse} assuming a common-envelope efficiency of $\alpha_{\mathrm{CE}}=0.15$, a Reimer's mass-loss parameter $\eta = 1.0$, and solar metallicity. However, the value that was adopted for the binding energy parameter $\lambda$ is not given in their paper. This parameter is of crucial importance in the \bse code as its value determines to what degree recombination energy is included in the energy budget of common envelope evolution.  

We therefore evolved the initial binary they proposed as a progenitor for NLTT\,12758 using \bse assuming a large range of values for $\lambda$. 
We used the same parameters as \citet{kawkaetal17-1} but different values for the fraction of recombination energy by varying $\alpha_{\mathrm{rec}}$ from 0 to 1 in steps of 0.1. Figure\,\ref{fig:NLT015} shows the results we obtained from this exercise. The two panels show the orbital separation of both stars after the first (left panel) and the second (right panel) common envelope phase as a function of the white dwarf mass. The assumed contributions from recombination energy are color coded and the dashed lines indicate the separation given for the corresponding stage in \citet{kawkaetal17-1}.  

Apparently, a high efficiency $\alpha_{\mathrm{rec}}$ between 0.7 and 0.8 was used by \citet{kawkaetal17-1}, i.e. they assumed that more than 70 per cent of the available recombination energy contributes to the envelope ejection process. Under these assumptions, during the first common envelope the binding energy is reduced by a factor of $\sim64$ and $\sim73$ respectively, i.e. $\lambda$ is 64\, and 73\, times larger for $\alpha_{\mathrm{rec}} = 0.7$ and $0.8$ with respect to $\lambda$ for $\alpha_{\mathrm{rec}} = 0$. This implies that while the orbital separation is reduced to $\sim1$ per cent during common envelope evolution for the model without recombination energy ($a_f/a_i \sim 0.01$), the simulations with $\alpha_{\mathrm{rec}} = 0.7$ and $0.8$ predict the separation after common envelope to be $\sim41$ and $\sim44$ per cent of the separation at the onset of common envelope evolution ($a_f/a_i \sim 0.41$ and $\sim 0.44$). 

We consider this assumption highly unlikely given that even simulations in favor of contributions from recombination energy conclude that at most 60 per cent of the initially available hydrogen recombination energy contribute to expelling the envelope \citep{ivanovaetal15-1}. Other works provide much stricter limits such as  \citet{sokeretal18-1} who found that at maximum 10 per cent of the available recombination energy can help expelling the envelope. 

The fact that we could not reproduce the evolution suggested by \citet{kawkaetal17-1} unless an unrealistic fraction of the recombination energy of the envelope contributes to the ejection (in both common envelope phases) might indicate that the authors perhaps misinterpreted the input parameter $\lambda$ of the {\bse} code. As noted by \citet{zorotovicetal14-1}, according to the comments written in the code by its developers one can read in the main routine (\textit{bse.f}) that the input parameter called lambda is "\textit{the binding energy factor for common envelope evolution (0.5)}". This leads many users to believe that $\lambda$ is fixed with the value they write in the input file.
However, the code has been updated after publication. In the README\_BSE file it says that the updates made after the publication of the {\bse} paper are documented in the header of the \textit{evolv2.f} routine. One of the updates documented there (since March 2001) says "\textit{The value of lambda used in calculations of the envelope binding energy for giants in common-envelope is now variable (see function in zfuncs)}". And if we take a look at the file called \textit{zfuncs.f} it is explained that the input parameter \textit{lambda} (that in this routine is called "\textit{fac}") is the fraction of the recombination/ionization energy used in the energy balance. Digging into the equations, which were fitted by Onno Pols and published in Robert Izzard's thesis (\citealt{Izzardtesis04}) and \citet[][their Appendix A]{claeysetal14}, one can confirm that the input parameter called \textit{lambda}, if positive, corresponds to the fraction of recombination energy that is assumed to contribute to the envelope ejection (called $\lambda_{\mathrm{ion}}$ in \citealt{claeysetal14} or $\alpha_{\mathrm{rec}}$ in \citealt{zorotovicetal14-1}).  
A fixed value for $\lambda$ can still be assumed in {\bse} by selecting a negative value. For example, if the input parameter is set to $-0.5$, then the common-envelope equations will use $\lambda=0.5$. If, on the other hand, the input parameter is set to $+0.75$, 75 per cent of the recombination energy is going to be used in the calculation of $\lambda$, which leads to an unrealistically small reduction in orbital separation during common envelope evolution.


\bsp	
\label{lastpage}
\end{document}